\documentclass[prl,twocolumn,groupedaddress]{revtex4}
\usepackage{graphicx,color}
\usepackage{amssymb}   % for math
\usepackage{amsmath}
\usepackage{epstopdf}
\usepackage{natbib}
\usepackage{hyperref}
\usepackage{bm}
\usepackage{color}

\usepackage{graphicx}
\graphicspath{{./img/}}

\let\mathbf=\boldsymbol
\def\bk{\textbf{k}}
\def\bso{\beta_\text{SO}}

\def\hc{H_{\text{c}2}}
\def\hp{H_\text{P}}
\begin{document}

\title{Proximity effect and Ising superconductivity in superconductor/transition metal dichalcogenide heterostructures}

\author{Ryohei Wakatsuki$^{1}$}\thanks{wakatsuki@appi.t.u-tokyo.ac.jp}
\author{Kam Tuen Law$^{2}$}\thanks{phlaw@ust.hk}
\affiliation{1. Department of Applied Physics, University of Tokyo, Hongo 7-3-1, 113-8656, Japan}
\affiliation{2. Department of Physics, Hong Kong University of Science and Technology, Clear Water Bay, Hong Kong, China}

\begin{abstract}
Recently, it was experimentally realized that 2D superconducting transition metal dichalcogenides (TMD) such as gated MoS$_2$ and monolayer NbSe$_2$ have in-plane upper critical magnetic fields much higher than the Pauli limit. This is due to the so-called Ising spin-orbit coupling (SOC) of TMD which pins the electron spins along the out-of-plane directions and protects the Cooper pairs from in-plane magnetic fields. However, many TMD materials with extremely large Ising SOC, in the order of a few hundred meV, are not superconducting. In this work, we show that TMD materials can induce strong Ising SOC on ordinary $s$-wave superconductors through proximity effect. By solving the self-consistent gap equation of the TMD/superconductor heterostructure, we found that the $H_{c2}$ of the $s$-wave superconductor can be strongly enhanced. Importantly, when the in-plane field is larger than the Pauli limit field and weaker than $H_{c2}$, the heterostructure becomes a nodal topological superconductor which supports Majorana flat bands.
\end{abstract}

\maketitle

\address{{\normalsize Department of Applied Physics, University of Tokyo, Hongo 7-3-1, 113-8656, Japan }}

\emph{\bf Introduction}--- Recently, the study of monolayer transition metal dichalcogenides (TMDs) has attracted much attention [\onlinecite{Wang,Xu1,LiuR}] because of their two dimensionality[\onlinecite{Mak0,Splendiani,Radisavljevic,Zhang,Bao}], valley degrees of freedom [\onlinecite{Cao,Zeng,Sallen,Mak1,Mak2,Gong}], Berry curvature physics [\onlinecite{Xiao1,Xiao2,Feng}], and strong spin--orbit coupling (SOC) [\onlinecite{Latzke,Yuan,Zhu,Feng,Kormanyos,Fang,Liu,Xiao2}].
Particularly, the strong SOC in the system is due to the breaking of in-plane mirror symmetry of the lattice structure and this strong SOC acts as an effective Zeeman field and polarizes the electron spins along the out-of-plane directions [\onlinecite{Latzke,Yuan,Zhu,Feng,Kormanyos,Fang,Liu,Xiao2}].
Moreover, time-reversal symmetry dictates that the effective Zeeman fields at opposite valleys have opposite directions.
Since the SOC pins electron spins along the $z$ directions only, we call it the Ising SOC to distinguish it from the Rashba SOC which originates from breaking out-of-plane mirror symmetry and pins electron spins to in-plane directions.
Importantly, in many TMD materials, the strength of the spin splitting due to Ising SOC near the $\pm K$ points is very strong [\onlinecite{Latzke,Yuan,Zhu,Feng,Kormanyos,Fang,Liu,Xiao2}].
For example, in WS$_2$, WSe$_2$, and WTe$_2$, the spin-orbit splitting exceeding 400 meV near the K-points of the top valence bands [\onlinecite{Zhu,Kormanyos,Liu,Fang}]. In the non-superconducting regime, several interesting phenomena related to this Ising SOC have been studied intensively in recent years [\onlinecite{Wang,Xu1,LiuR}].

On the other hand, the experimental study of the superconducting properties of mono and few layer TMDs have only started recently but interesting phenomena such as Ising superconductivity [\onlinecite{Lu,Xi1,Saito}] and a quantum metal phase have been found [\onlinecite{Tsen}]. Particularly, superconducting properties of gated MoS$_2$ and few layer NbSe$_2$ have been investigated [\onlinecite{Ye,Taniguchi,Shi,Yokoya,Rahn,Xi2,Ugeda,Johannes,Lu,Xi1,Saito, Tsen,Bawden}]. Unlike conventional $s$-wave superconductors in which superconductivity can be destroyed by magnetic fields stronger than the Pauli limit field ($H_\text{P}=1.86T_\text{c}$) through Zeeman effects [\onlinecite{Clogston,Chandrasekhar}], the in-plane upper critical fields $H_{\text{c}2}$ of mono and few layer TMD materials can be several times stronger than the Pauli limit [\onlinecite{Lu,Xi1,Saito}].
As explained recently, this robustness against the in-plane magnetic field can be explained by their two dimensionality and the Ising SOC  [\onlinecite{Lu,Xi1,Saito, Wenyu}]. First of all, the two dimensional nature of the atomically thin films strongly suppresses the orbital effect.
Secondly, the strong pinning of the Cooper pair spins along the out-of-plane directions makes them robust against the Zeeman effects of in-plane magnetic fields.

Indeed, the Ising SOC splittings at the Fermi energy in gated MoS$_2$ and monolayer NbSe$_2$ are about 10meV and 30meV respectively. This is already sufficient to enhance $H_{\text{c}2}$ to several times higher than $H_\text{P}$ [\onlinecite{Lu,Xi1,Saito, Wenyu}]. As mentioned above, many TMD materials such as W based TMDs have exceedingly large Ising SOC but they are not superconducting. The question is: Can we make use of these materials to create superconductors with strongly enhanced $H_{\text{c}2}$?

In this work, we show that when a conventional $s$-wave superconductor is in proximity to a TMD material, the TMD material can induce Ising SOC on the $s$-wave superconductor similar to the case of graphene on TMD materials [\onlinecite{Gmitra}]. By solving the self-consistent gap equations, we show that even though the $T_\text{c}$ of the superconductor can be reduced due to inverse proximity effect, $H_{\text{c}2}$ can be strongly enhanced to several times above the Pauli limit. We demonstrate that this TMD/superconductor heterostructure can be used to create a nodal topological superconductor with a large number of zero energy Majorana modes residing on the edges of the superconductor.

The rest of the paper is organized as follows.
First, we explain how Ising SOC can be induced on a conventional $s$-wave superconductor by a TMD material through proximity effect.
Second, by solving the self-consistent gap equation of the heterostructure, we show that the $H_{\text{c}2}$ of the superconductor is strongly enhanced.  Third, we point out that, when the applied in-plane field is higher than the Pauli limit and lower than $H_{\text{c}2}$, the system is in a nodal topological phase which supports Majorana flat bands. %We believe that this work will open a new way for exploring the applications of TMD with extremely strong Ising SOC and to study Ising superconductivity.

\begin{figure}[t]
\centerline{\includegraphics[width=0.5\textwidth]{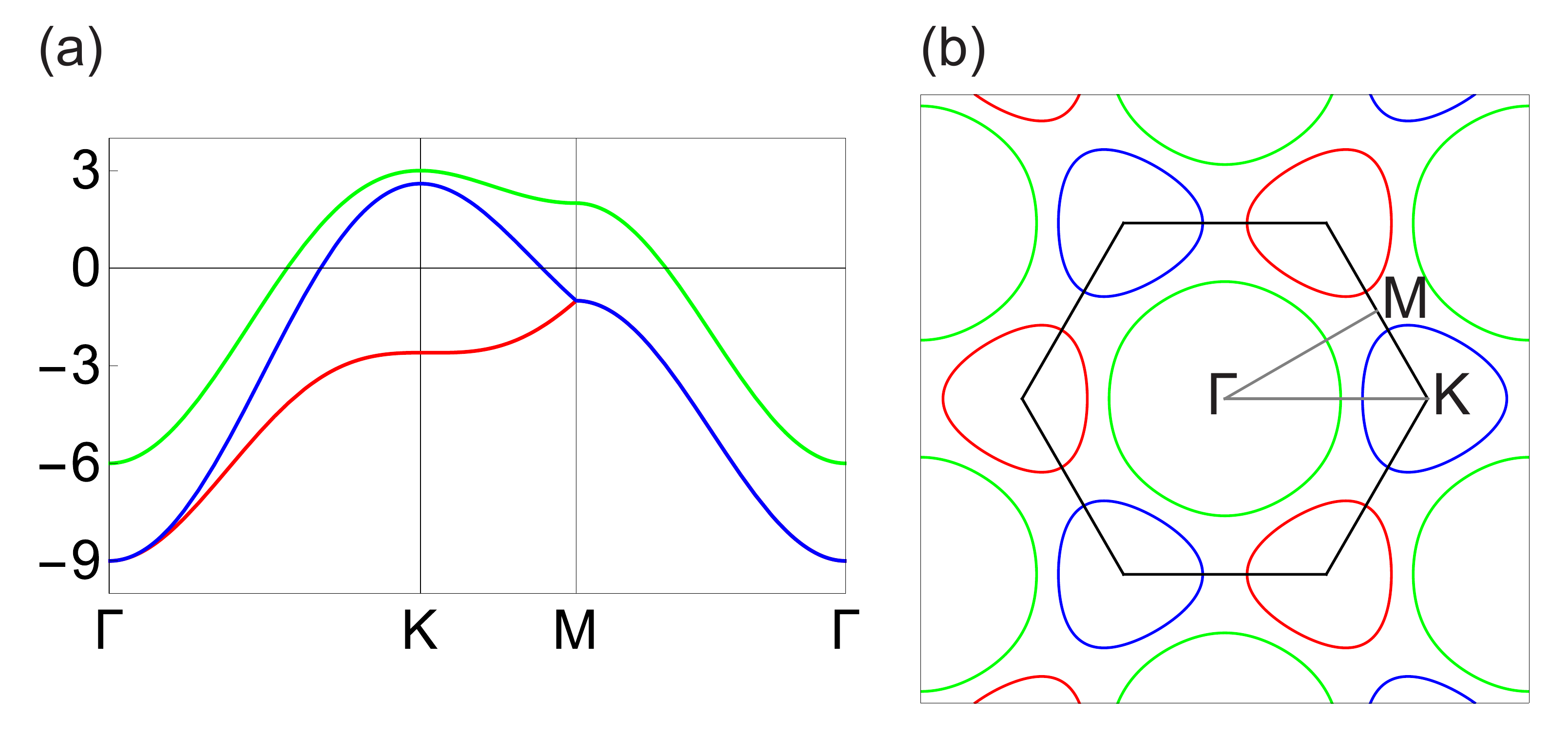}}
\caption{
(a) The band structures of the superconductor (green line) and TMD (red and blue lines). Parameters of $H_\text{TMD}$ and $H_\text{BdG}$ are: $t_\text{SC}=t_\text{TMD}=-1.0, \mu_\text{SC}=0.0, \mu_\text{TMD}=3.0$, and $\bso=0.5$. As $H_\text{TMD}$ preserves spin in the out-of-plane directions, the red and blue bands have opposite spins.
(b) The Fermi surfaces of the superconductor and TMD. The black line denotes the first Brillouin zone boundary. The red and blue Fermi pockets have opposite spins due to Ising SOC.
}
\label{Fig:band}
\end{figure}

{\bf Model Hamiltonian}---
We start with a simple two-band model which captures the symmetry and describes the top valence bands of a monolayer TMD material.
In the basis of $[ c_{k \uparrow}, c_{k \downarrow} ]$, the effective Hamiltonian of the top valence bands near the $\pm K$ points, is [\onlinecite{Noah}]
\begin{equation}
H_\text{TMD}\left(\bk\right)=\left(2t_\text{TMD}C\left(\bk\right)-\mu_\text{TMD}\right)\sigma_0+2\bso S\left(\bk\right)\sigma_z , \label{Eq:HamiltonianTMD}
\end{equation}
where $\sigma_z$ is the Pauli matrix for the spin space, and $C\left(\bk\right)$ and $S\left(\bk\right)$ are defined as
\begin{equation}
\begin{array}{ll}
C\left(\bk\right)&=\sum_{i=1,2,3} \cos\left(\bk \cdot \mathbf{R}_i\right), \\
S\left(\bk\right)&=\sum_{i=1,2,3} \sin\left(\bk \cdot \mathbf{R}_i\right),  \\
R_1&=\left(a,0,0\right),  \\
R_2&=\left(-\frac{a}{2}, +\frac{\sqrt{3}a}{2}\right),  \\
R_3&=\left(-\frac{a}{2}, -\frac{\sqrt{3}a}{2}\right).
\end{array}
\end{equation}
This Hamiltonian preserves $C_{3v}$ symmetry, out-of-plane mirror symmetry and time-reversal symmetry.
However, the in-plane mirror symmetry is broken and results in the second term of Eq.\ref{Eq:HamiltonianTMD} which is the Ising SOC term.
$S\left(\bk\right)$ is an odd function of $\bk$ and this Ising SOC pins the electron spins in the opposite out-of-plane directions for electrons with opposite momenta. In the presence of an in-plane magnetic field in the $x$-direction, a term $V_x\sigma_x=2\mu_BH_x\sigma_x$ is added to the Hamiltonian where $\mu_B$ is the Bohr magneton. It is noted that the spin splitting at the $\pm K$ points is $3\sqrt{3}\beta_\text{SO}$ in our model. The band structure of the Hamiltonian is depicted in Fig.\ref{Fig:band}a (blue and red curves). The Fermi surfaces near the K points are depicted in Fig.\ref{Fig:band}b (blue and red curves). Due to the large SOC of the TMD, we can assume only a single spin band crosses the Fermi energy as in the cases of some W-based TMDs. As depicted in Fig.\ref{Fig:band}b, the Fermi surfaces at opposite K points are depicted in different colors to emphasize the fact that the spin polarizations at opposite valleys have opposite signs.

To study the heterostructure formed by a TMD and a conventional $s$-wave superconductor, we assume that the $s$-wave superconductor has a simple Fermi surface enclosing the $\Gamma$ point as depicted in Fig.\ref{Fig:band}a and Fig.\ref{Fig:band}b (the green lines). %The effects of more complicated Fermi surfaces will be discussed at the end of the manuscript. 
In the basis of $[s_{k \uparrow}, s_{k \downarrow}, s_{-k \uparrow}^{\dagger}, s_{-k \downarrow}^{\dagger}]$, the BdG Hamiltonian of the superconductor can be written as:
\begin{equation}
H_\text{BdG} = \left(2t_\text{SC}C\left(\bk\right)-\mu_\text{SC}\right)\sigma_0\tau_z+\Delta \sigma_y \tau_y +V_x \sigma_x \tau_z. \label{Eq:HamiltonianBdG}
\end{equation}
Here, $\tau_i$ are Pauli matrices acting on particle-hole basis and the $V_x$ is caused by an in-plane magnetic field in the $x$-direction. Then, we consider the coupling of the TMD and the superconductor by turning on the spin-independent hopping terms $\Gamma c_{k}^{\dagger}s_{k} + h.c. $. With the coupling terms, the Green's function of the superconductor is
\begin{equation}
\begin{array}{ll}
G_\text{BdG}&=\left(i\omega-H_\text{BdG}-\Sigma\right)^{-1}, \\
\Sigma&=\Gamma^2\left(i\omega-H_\text{TMD}\right)^{-1}.
\end{array}
\end{equation}
Here, $\Sigma$ is the self-energy contribution to the Green function due to the TMD. At $\omega = 0$, the effective Hamiltonian of the superconductor becomes $H_{\text{eff}}=H_{\text{BdG}} + \Sigma (\omega=0)$. It is straight-forward to show that $H_{\text{eff}}$ acquires an Ising SOC term $2\tilde{\beta}_{\text{so}}\tilde{S}({\bk})\sigma_z$ due to the coupling between the superconductor and the TMD, where 
\begin{equation}
2 \tilde{\beta}_\text{SO} \tilde{S}\left(\bk\right) \equiv \Gamma^2 \frac{-2\bso S\left(\bk\right)}{\left(2t_\text{TMD}C\left(\bk\right)-\mu_\text{TMD}\right)^2+\left(2\bso S\left(\bk\right)\right)^2}.
\end{equation}
%We expect that this Ising SOC term can strongly enhances the in-plane $H_{c2}$ of the SC as we demonstrate in the following sections.
%This Ising SOC term is large around the Fermi surface of the TMD.
%Therefore, to induce large Ising SOC on the SC, the Fermi surfaces of the TMD and the SC should be sufficiently close to each other. 
\begin{figure}[t]
\centerline{\includegraphics[width=0.5\textwidth]{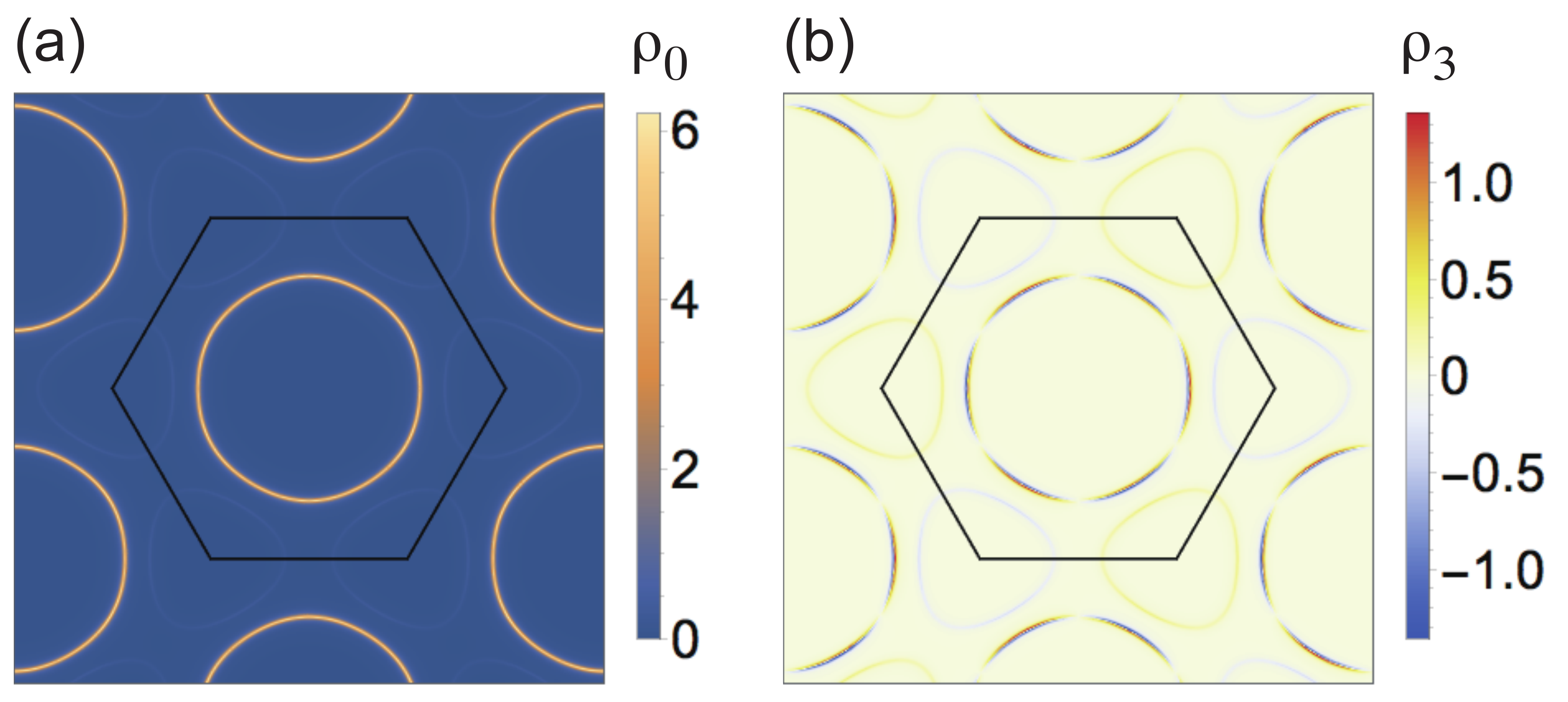}}
\caption{
(a) The density of states $\rho_0$ for the superconductor film in the normal state.
(b) The spin density of states $\rho_3$ for the superconductor film in the normal state. $\rho_3$ is opposite at opposite k points and vanishes along $\Gamma$-$M$ lines.
}
\label{Fig:dos}
\end{figure}

To study the effect of the induced Ising SOC terms, we first study the normal state property of the heterostructure by setting $\Delta = 0$.
By assuming the coupling strength to be $\Gamma=0.5$ and using the retarded Green's function $G^{\text{R}}_{\text{BdG}}$, we can calculate the spin up and spin down density of states $\rho_\uparrow$ and $\rho_\downarrow$ where:
\begin{equation}
\rho_{\sigma} \left(\bk,\omega\right) = -\frac{1}{\pi}\mathrm{Im}[G^{\text{R}}_{\text{BdG}} \left(\bk,\omega\right)]_{\sigma,\sigma}.
\end{equation}
With $\rho_\uparrow$ and $\rho_\downarrow$, we can define the total density of states $\rho_0=\rho_\uparrow+\rho_\downarrow$ and the net spin density of state $\rho_3=\rho_\uparrow-\rho_\downarrow$ at the Fermi surface.

The density of states $\rho_0$ and $\rho_3$ are illustrated in Fig.\ref{Fig:dos}a and Fig.\ref{Fig:dos}b respectively. It is important to note that, at the Fermi energy, the superconductor acquires non-zero net spin density of state due to the effective Ising SOC terms induced on the superconductor by the TMD. The maximal amplitude of $\rho_3$ is in the order of the $\rho_0$ which indicates that Ising SOC can be effectively induced on the superconductor film even though there is momentum mismatch between the superconductor and the TMD Fermi surfaces. %It is also interesting to note that the SOC vanishes along the $\Gamma-M$ lines of the Brillouin zone due to the point group symmetries of the system. Therefore, the net spin density vanishes along the $\Gamma-M$ lines as well.

Due to time-reversal symmetry, the net spin polarization $\rho_3$ has opposite signs at opposite K-points as shown in Fig.\ref{Fig:dos}b. It is interesting to note that the Ising SOC vanishes along the $\Gamma$-$M$ lines in the Brillouin zone due to the point group symmetries of the TMD material. Along the $\Gamma$-$M$ line, the spin-up and spin-down states are degenerate. However, this degeneracy can be lifted by an in-plane magnetic field. As we will show below, this property is important in making the heterostructure a nodal topological superconductor with Majorana flat bands.

\begin{figure}[t]
\centerline{\includegraphics[width=0.5\textwidth]{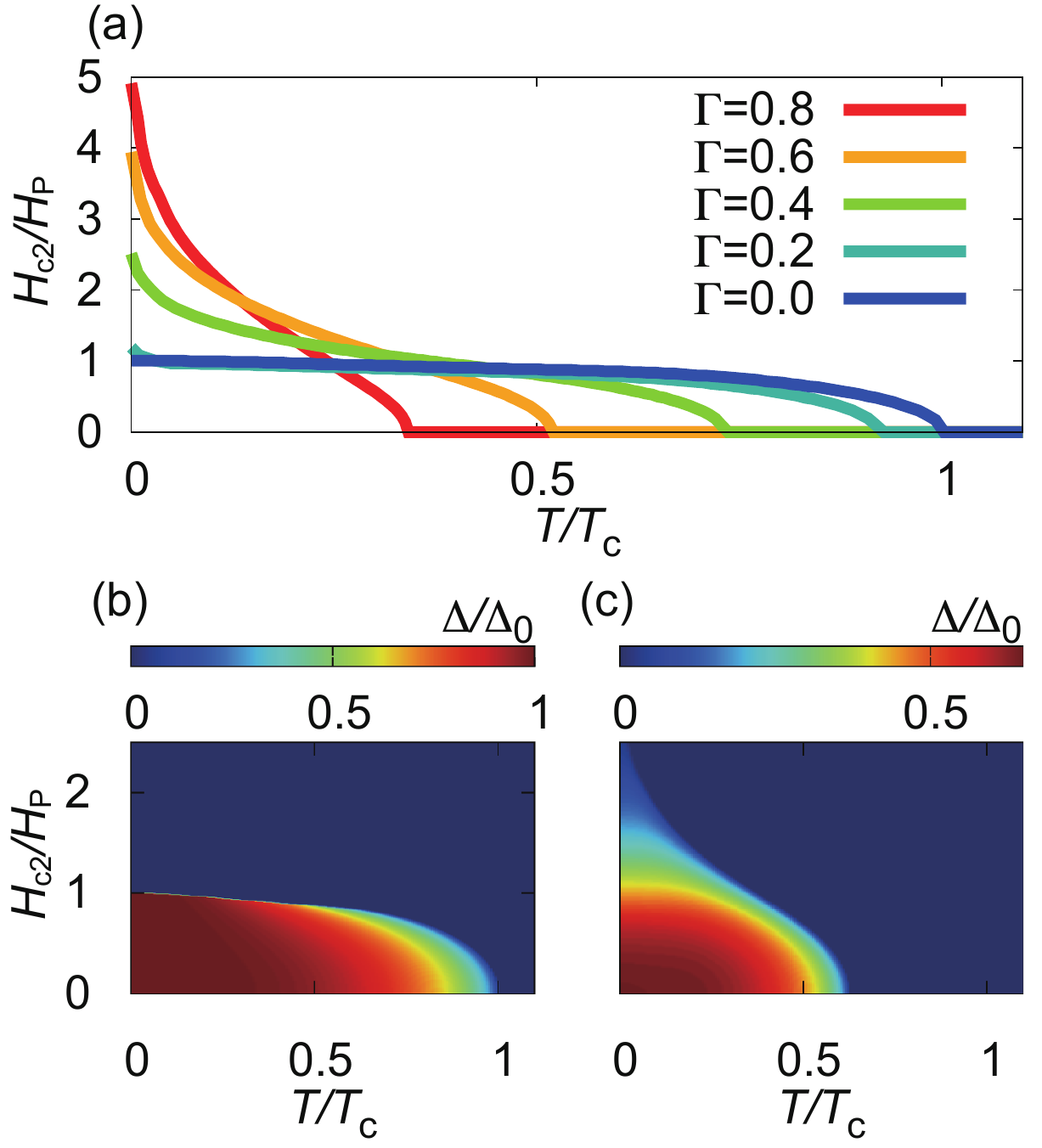}}
\caption{
(a) $H_{c2}$ as a function of temperature for various $\Gamma$. $H_{c2}$ can go beyond the Pauli limit due to the induced Ising SOC.
(b) $\Delta/\Delta_0$ for an isolated layer of $s$-wave superconductor. The superconductor to metal phase transition driven by magnetic field at low temperatures is first order. (c) $\Delta/\Delta_0$ with $\Gamma = 0.5$ for the superconductor/TMD heterostructure. The phase transition driven by magnetic field is second order at low temperatures.
}
\label{Fig:hc}
\end{figure}

{\bf Enhancement of $H_{c2}$}--- 
In this section, we calculate the superconducting phase diagram of the superconductor/TMD heterostructure by solving the self-consistent mean field equation. As shown in Fig.\ref{Fig:dos}b, the electron spins are pinned to the out-of-plane directions by Ising SOC. Therefore, in-plane magnetic fields are not effective in aligning the electron spins to the in-plane direction and we expect that the in-plane $H_{c2}$ is strongly enhanced as in the case of intrinsically superconducting thin films [\onlinecite{Lu,Xi1,Saito}].
To perform a self-consistent calculation, we replace the pairing terms in the $H_{\text{BdG}}$ of the superconductor in Eq.\ref{Eq:HamiltonianBdG} by interaction terms of the form $ -v\sum_{{k} {q}} s_{{k} \uparrow}^{\dagger}s_{-{k} \downarrow}^{\dagger}s_{{q} \uparrow}s_{-{q}  \downarrow} $ where $v$ a positive number. Together with the $H_{\text{TMD}}$ in Eq.\ref{Eq:HamiltonianTMD} and the coupling between the TMD and the superconductor, the standard self-consistent gap equations of the heterostructure is:
\begin{equation}
\Delta= v \sum_{k}\langle s_{k \uparrow}s_{k \downarrow} \rangle.
\end{equation}
From the gap equation, we can calculate the critical field $H_{c2}$ at which the pairing potential $\Delta$ vanishes as a function of temperature. 
The resulting $H_{c2}$ for various coupling strength $\Gamma$ is illustrated in Fig.\ref{Fig:hc}.
The temperature and the magnetic field are normalized by the critical temperature $T_c$ at zero field and the Pauli limit $\hp$ at zero temperature for the monolayer superconductor.
The line with $\Gamma=0$ depicts the case where the superconductor is decoupled from the TMD, and $\hc$ is limited by the Pauli limit $\hp$ as expected.
It is important to note that, at zero magnetic field, the superconducting transition temperature $T_c$ decreases as $\Gamma$ increases due to inverse proximity effect. However, even though $T_c$ is decreased, the in-plane $H_{c2}$ is strongly enhanced and can go beyond the Pauli limit field.
This enhancement of $H_{c2}$ is due to the induced SOC from the TMD [\onlinecite{Lu, Xi1, Saito}].

To understand the evolution of $\Delta$ under in-plane magnetic fields, we calculate $\Delta/\Delta_0$ as a function of magnetic field and temperature where $\Delta_0$ is the pairing potential at zero temperature and zero magnetic field. The results are depicted in Fig.\ref{Fig:hc}b and Fig.\ref{Fig:hc}c. At low temperatures, magnetic field can drive the superconductor to metal through first order phase transition. However, when the superconducting thin film is coupled to the TMD, the Ising SOC changes the first order transition to second order and $H_{c2}$ is strongly enhanced.
Hence, we predict that the $H_{c2}$ of a conventional $s$-wave superconductor can be enhanced when it is coupled to a TMD thin film. This is one of the important findings of this work.

{\bf Nodal topological superconductor}---
In the above section, we solved the self-consistent gap equation of the heterostructure and found that the $H_{c2}$ of the heterostructure can go beyond the Pauli limit. The next question is, what happens when the applied magnetic field is beyond the Pauli limit but smaller than $H_{c2}$? In this section, we show that the applied in-plane magnetic field can drive the heterostructure into a nodal topological superconducting phase through a topological phase transition similar to the case of intrinsic Ising superconductors [\onlinecite{Wenyu}].

\begin{figure}[t]
\centerline{\includegraphics[width=0.5\textwidth]{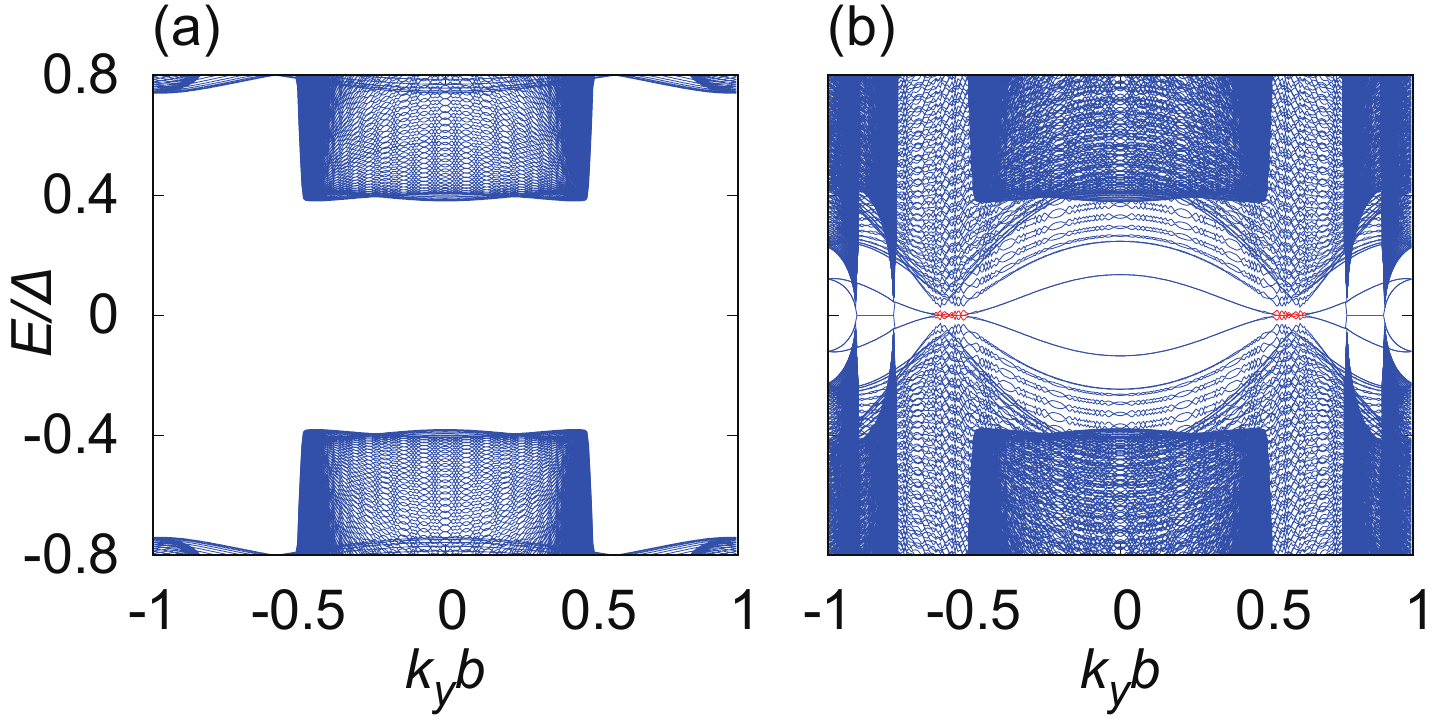}}
\caption{
Energy spectrums of a TMD/superconductor strip with armchair edges. (a)  When the applied magnetic field  is zero ($V_x=0$), the pairing phase is fully gapped. (b) When the applied magnetic field is larger than the pairing gap ($V_x=4\Delta$), there are nodal points in the energy spectrum. The nodal points are connected by Majorana flat bands. The Majorana flat bands are associated with a large number of Majorana modes residing on the armchair edge of the sample. Here, $b=\frac{\sqrt{3}a}{\pi}$. }
\label{Fig:mfb}
\end{figure}

To further understand the effect of the in-plane magnetic field, we plot the energy spectrum of a strip of TMD/superconductor heterostructure which has open boundary conditions in the $x$-direction and periodic boundary conditions in the $y$-direction. When the in-plane magnetic field is smaller than the Pauli limit, the superconductor remains fully gapped as shown in Fig.\ref{Fig:mfb}a. When the applied in-plane magnetic field is larger than the pairing gap energy, the bulk superconducting gap is closed and nodal points emerge as shown in Fig.\ref{Fig:mfb}b. Importantly, the nodal points are connected by Majorana flat bands, similar to the case of Weyl points being connected by surface Fermi arcs in Weyl semimetals.

To understand the origin of the nodal points and the Majorana flat band, we integrate out the TMD and derive the effective Hamiltonian of the superconducting thin film.
In the basis of $[s_{k \uparrow}, s_{k \downarrow}, s_{-k \uparrow}^{\dagger}, s_{-k \downarrow}^{\dagger}]$ and in the presence of an in-plane magnetic field $V_{x}$, the effective Hamiltonian of the superconductor can be written as:
\begin{equation}
\begin{array}{ll}
H_\text{eff} (k_x, k_y)  = & \left(2\tilde{t}_\text{SC}C\left(\bk\right)-\tilde{\mu}_\text{SC}\right)\sigma_0\tau_z+ 2 \tilde{\beta}_\text{SO} \tilde{S}\left(\bk\right)\sigma_z \tau_0 \\  & + V_{x} \sigma_x \tau_z + \Delta \sigma_y \tau_y. \label{Eq:Heff}
\end{array}
\end{equation}

Even though time-reversal symmetry is broken by the magnetic field $V_x$, the $H_\text{eff}$ satisfy a 1D time-reversal like symmetry $ T_{1D}H_\text{eff} (k_x, k_y) T_{1D}^{-1} = H_\text{eff} (-k_x, k_y)$ and a 1D particle-hole like symmetry $ P_{1D}H_\text{eff} (k_x, k_y) P_{1D}^{-1} = -H_\text{eff} (-k_x, k_y)$, where $T = \sigma_x \tau_z K$ ($T^2=1$) and $P = \sigma_0 \tau_x K$ ($P^2=1$). Here, $K$ is the complex conjugate operator. As a result, $H_\text{eff}$ satisfies a chiral symmetry $C H_\text{eff}(k_x,k_y) C^{-1} = -H_\text{eff}(k_x,k_y)$ where $C = \sigma_x \tau_y$. Therefore, for a fixed $k_y$, the Hamiltonian $H_\text{eff}(k_x,k_y)=H_{k_y}(k_x)$ is in the BDI class according to the Altland-Zirnbauer classification [\onlinecite{Altland,Schnyder,Kitaev,Ryu}]. 

In the absence of external magnetic fields, $H_{k_y} (k_x)$ is fully gapped and the system is topologically trivial for all $k_y$. However, when the in-plane magnetic field satisfies the condition $V_x > \Delta$, the system becomes topologically non-trivial for a range of $k_y$. At the critical values of $k_y$ which separates the topologically trivial and non-trivial regimes, the bulk gap has to close due to the topological phase transition and the nodal points emerge. The nodal points are connected by Majorana flat bands. As shown previously, these Majorana modes can induced selective equal spin Andreev reflections which can be used to generate spin polarized currents [\onlinecite{James, Noah2}]. We believe the TMD/superconductor heterostructure can open a way to realize nodal topological superconductors and may have applications in superconducting spintronics [\onlinecite{Linder, Eschrig}].

{\bf Conclusion and Discussion}--- It is important to note that the proposed superconductor/TMD heterostructure has very important advantages for studying Ising superconductivity. First, one can use many non-superconducting TMD materials with strong SOC to study Ising superconductivity. 
Second, one can study the effects of Ising SOC on superconducting thin films with higher $T_\text{c}$ and  even with unconventional pairings. The resulting heterostructure with strong Ising SOC can be used to realize Majorana fermions at higher temperatures and have applications in superconducting spintronics.

{\bf Acknowledgements}--- RW was supported by Grant-in-Aid for JSPS Fellows. KTL thanks the support of HKRGC and Croucher Foundation through HKUST3/CRF/13G, 602813, 605512, 16303014 and Croucher Innovation Grant.


\begin{thebibliography}{99}

%<TMD review>
\bibitem{Wang}
Q. H. Wang, K. K.-Zadeh, A. Kis, J. N. Coleman, and M. S. Strano,
Nat. Nanotechnol. \textbf{7}, 699--712 (2012).

\bibitem{Xu1}
X. Xu, W. Yao, D. Xiao, and T. F. Heinz,
Nat. Phys. \textbf{10}, 343--350 (2014).

\bibitem{LiuR}
G.-B. Liu, D. Xiao, Y. Yao, X. Xu, and W. Yao,
Chem. Soc. Rev. \textbf{44}, 2643 (2015).

%<TMD>
\bibitem{Mak0}
K. F. Mak, C. Lee, J. Hone, J. Shan, and T. F. Heinz,
Phys. Rev. Lett. \textbf{105}, 136805 (2010).

\bibitem{Splendiani}
A. Splendiani, L. Sun, Y. Zhang, T. Li, J. Kim, C.-Y. Chim, G. Galli, and F. Wang,
Nano Lett. \textbf{10}, 1271--1275 (2010).

\bibitem{Radisavljevic}
B. Radisavljevic, A. Radenovic, J. Brivio, V. Giacometti, and A. Kis,
Nat. Nanotechnol. \textbf{6}, 147--150 (2011).

\bibitem{Zhang}
Y. Zhang, J. Ye, Y. Matsuhashi, and Y. Iwasa,
Nano Lett. \textbf{12}, 1136--1140 (2012).

\bibitem{Bao}
W. Bao, X. Cai, D. Kim, K. Sridhara, and M. S. Fuhrer,
Appl. Phys. Lett. \textbf{102}, 042104 (2013).


%<Valley>
\bibitem{Cao}
T. Cao, G. Wang, W. Han, H. Ye, C. Zhu, J. Shi, Q. Niu, P. Tan, E. Wang, B. Liu, and J. Feng,
Nat. Commun. \textbf{3}, 887 (2012).

\bibitem{Zeng}
H. Zeng, J. Dai, W. Yao, D. Xiao, X. Cui,
Nat. Nanotechnol. \textbf{7}, 490--493 (2012).

\bibitem{Sallen}
G. Sallen, L. Bouet, X. Marie, G. Wang, C. R. Zhu, W. P. Han, Y. Lu, P. H. tan, T. Amad, B. L. Liu, and B. Urbaszek,
Phys. Rev. B \textbf{86}, 081301(R) (2012).

\bibitem{Mak1}
K. F. Mak, K. He, J. Shan, and T. F. Heinz,
Nat. Nanotechnol. \textbf{7}, 494--498 (2012).

\bibitem{Gong}
Z. Gong, G.-B. Liu, H. Yu, D. Xiao, X. Cui, X. Xu, and W. Yao,
Nat. Commun. \textbf{4}, 2053 (2013).

\bibitem{Mak2}
K. F. Mak, K. L. McGill, J. Park, and P. L. McEuen,
Science \textbf{344}, 1489--1492 (2014).

%\bibitem{Cai}
%T. Cai, S. A. Yang, X. Li, F. Zhang, J. Shi, W. Yao, and Q. Niu,
%Phys. Rev. B \textbf{88}, 115140 (2013).

%\bibitem{Li}
%Y. Li, J. Ludwig, T. Low, A. Chernikov, X. Cui, G. Arefe, Y. D. Kim, A. M. van der Zande, A. Rigosi, H. M. Hill, S. H. Kim, J. Hone, Z. Li, D. Smirnov, and T. F. Heinz,
%Phys. Rev. Lett. \textbf{113}, 226804 (2014).
%
%\bibitem{MacNeill}
%D. MacNeill, C. Heikes, K. F. Mak, Z. Anderson, A. Korm\'{a}nyos, V. Z\'{o}lyomi, J. Park, and D. C. Ralph,
%Phys. Rev. Lett. \textbf{114}, 037401 (2015).
%
%\bibitem{Srivastava}
%A. Srivastava, M. Sidler, A. V. Allain, D. S. Lembke, A. Kis, and A. Imamo\u{g}lu,
%Nat. Phys. \textbf{11}, 141--147 (2015).
%
%\bibitem{Aivazian}
%G. Aivazian, Z. Gong, A. M. Jones, R.-L. Chu, J. Yan, D. G. Mandrus, C. Zhang, D. Cobden, W. Yao, and X. Xu,
%Nat. Phys. \textbf{11}, 148--152 (2015).

%<Berry phase>
\bibitem{Xiao1}
D. Xiao, W. Yao, and Q. Niu,
Phys. Rev. Lett. \textbf{99}, 236809 (2007).

\bibitem{Xiao2}
D. Xiao, G.-B. Liu, W. Feng, X. Xu, and W. Yao,
Phys. Rev. Lett. \textbf{108}, 196802 (2012).

\bibitem{Feng}
W. Feng, Y. Yao, W. Zhu, J. Zhou, W. Yao, and D. Xiao,
Phys. Rev. B \textbf{86}, 165108 (2012).

%<First principle>
\bibitem{Zhu}
Z. Y. Zhu, Y. C. Cheng, and U. Schwingenshl\"{o}gl,
Phys. Rev. B \textbf{84}, 153402 (2011).

\bibitem{Kormanyos}
A. Korm\'{a}nyos, V. Z\'{o}lyomi, N. D. Drummond, P. Rakyta, G. Burkard, and V. I. Fal'ko,
Phys. Rev. B \textbf{88}, 045416 (2013).

\bibitem{Liu}
G.-B. Liu, W.-Y. Shan, Y. Yao, W. Yao, and D. Xiao,
Phys. Rev. B \textbf{88}, 085433 (2013).

\bibitem{Fang}
S. Fang, R. K. Defo, S. N. Shirodkar, S. Lieu, G. A. Tritsaris, and E. Kaxiras,
Phys. Rev. B \textbf{92}, 205108 (2015).

%<Zeeman>
\bibitem{Yuan}
H. Yuan, M. S. Bahramy, K. Morimoto, S. Wu, K. Nomura, B.-J. Yang, H. Shimotani, R. Suzuki, M. Toh, C. Kloc, X. Xu, R. Arita, N. Nagaosa, and Y. Iwasa,
Nat. Phys. \textbf{9}, 563--569 (2013).

\bibitem{Latzke}
D. W. Latzke, W. Zhang, A. Suslu, T.-R. Chang, H. Lin, H.-T. Jeng, S. Tongay, J. Wu, A. Bansil, and A. Lanzara,
Phys. Rev. B, \textbf{91}, 235202 (2015).


%<Ising SC exp.>
\bibitem{Lu}
J. M. Lu, O. Zeliuk, I. Leermakers, N. F. Q. Yuan, U. Zeitler, K. T. Law, and J. T. Ye,
Science \textbf{350}, 1353--1357 (2015).

\bibitem{Xi1}
X. Xi, Ze Wang, W. Zhao, J.-H. Park, K. T. Law, H. Berger, L.  Forr\'{o}, J. Shan, and K. F. Mak,
Nature Physics (2015), DOI:10.1038/nphys3538.

\bibitem{Saito}
Y. Saito, Y. Nakamura, M. S. Bahramy, Y. Kohama, Y. Kasahara, Y. Nakagawa, M. Onga, M. Tokunaga, T. Nojima, Y. Yanase, and Y. Iwasa,
Nature Physics (2015), DOI:10.1038/nphys3580.

%<Quantum Metal>
\bibitem{Tsen}
A. W. Tsen, B. Hunt, Y. D. Kim, Z. J. Yuan, S. Jia, R. J. Cava, J. Hone, P. Kim, C. R. Dean, and A. N. Pasupathy,
Nat. Phys. (2015), DOI:10.1038/nphys3579

%<superconducting TMD>
\bibitem{Ye}
J. T. Ye, Y. J. Zhang, R. Akashi, M. S. Bahramy, R. Arita, and Y. Iwasa,
Science \textbf{338}, 1193--1196 (2012).

\bibitem{Taniguchi}
K. Taniguchi, A. Matsumoto, H. Shimotani, and H. Takagi,
Appl. Phys. Lett. \textbf{101}, 042603 (2012).

\bibitem{Shi}
W. Shi, J. Ye, Y. Zhang, R. Suzuki, M. Yoshida, J. Miyazaki, N. Inoue, Y. Saito, and Y. Iwasa,
Sci. Rep. \textbf{5}, 12534 (2015).

%<NbSe2>
\bibitem{Yokoya}
T. Yokoya, T. Kiss, A. Chainani, S. Shin, M. Nohara, and H. Takagi,
Science \textbf{294}, 2518--2520 (2001).

\bibitem{Rahn}
D. J. Rahn, S. Hellmann, M. Kall\"{a}ne, C. Sohrt, T. K. Kim, L. Kipp, and K. Rossnagel,
Phys. Rev. B \textbf{85}, 224532 (2012).

\bibitem{Xi2}
X. Xi, L. Zhao, Z. Wang, H. Berger, L. Forr\'{o}, J. Shan, and K. F. Mak,
Nat. Nanotechnol. \textbf{10}, 765--769 (2015).

\bibitem{Ugeda}
M. M. Ugeda, A. Bradley, Y. Zhang, S. Onishi, Y. Chen, W. Ruan, C. Ojeda-Aristizabal, H. Ryu, M. T. Edmonds, H.-Z. Tsai, A. Riss, S.-K. Mo, D. Lee, A. Zettl, Z. Hussain, Z.-X. Shen, and M. F. Crommie,
arXiv:1506.08460.

\bibitem{Johannes}
M. D. Johannes, I. I. Mazin, and C. A. Howells,
Phys. Rev. B \textbf{73}, 205102 (2006).

\bibitem{Bawden}
L. Bawden et al. arXiv:1603.05207.

%<Pauli limit>
\bibitem{Clogston}
A. M. Clogston,
Phys. Rev. Lett. \textbf{9}, 266 (1962).

\bibitem{Chandrasekhar}
B. S. Chandrasekhar,
Appl. Phys. Lett. \textbf{1}, 7 (1962).

%%<upper critical field>
%\bibitem{Frigeri1}
%P. A. Frigeri, D. F. Agterberg, A. Koga, and M. Sigrist,
%Phys. Rev. Lett. \textbf{92} ,097001 (2004).
%
%\bibitem{Frigeri2}
%P. A. Frigeri, D. F. Agterberg, and M. Sigrist,
%New. J. Phys. \textbf{6}, 115 (2004).
%
%%<Majorana review>
%\bibitem{Alicea}
%J. Alicea,
%Rep. Prog. Phys. \textbf{75}, 076501 (2012).
%
%\bibitem{Beenakker}
%C. W. J. Beenakker
%Annu. Rev. Con. Mat. Phys. \textbf{4}, 113 (2013).

\bibitem{Wenyu}
W-Y He, B. T. Zhou, J. J. He, T. Zhang and K. T. Law
arXiv:1604.02867.

%<Proximity>
\bibitem{Gmitra}
M. Gmitra and J. Fabian,
Phys. Rev. B, \textbf{92}, 155403 (2015).

%<TMD Majorana>
\bibitem{Noah}
N. F. Q. Yuan, K. F. Mak, and K. T. Law,
Phys. Rev. Lett. \textbf{113}, 097001 (2014).

%\bibitem{Zhou}
%T. Zhou, H.-L. Jiang, N. F. Q. Yuan, and K. T. Law,
%arXiv:1510.06289.

%\bibitem{Chu}
%R.-L. Chu, G.-B. Liu, W. Yao, X. Xu, D. Xiao, and C. Zhang,
%Phys. Rev. B \textbf{89}, 155317 (2014).
%
%\bibitem{Xu2}
%G. Xu, J. Wang, B. Yan, and X.-L. Qi,
%Phys. Rev. B \textbf{90}, 100505(R) (2014).

%%<2D SOC SC>
%\bibitem{Yip}
%S. K. Yip,
%Phys. Rev. B \textbf{65}, 144508 (2002).
%
%\bibitem{Gorkov}
%L. P. Gor'kov and E. I. Rashba,
%Phys. Rev. Lett. \textbf{87}, 3 (2001).

%<Classification>

\bibitem{Altland}
A. Altland and M. R. Zirnbauer,
Phys. Rev. B, \textbf{55}, 1142 (1997).

\bibitem{Schnyder}
A. P. Schnyder, S. Ryu, A. Furusaki, and A. W. W. Ludwig,
Phys. Rev. B, \textbf{78}, 195125 (2008).

\bibitem{Kitaev}
A. Kitaev,
AIP Conf. Proc. \textbf{1134}, 22 (2009).

\bibitem{Ryu}
S. Ryu, A. P. Schnyder, A. Furusaki, and A. W. W. Ludwig,
New. J. Phys. \textbf{12}, 065010 (2010).

\bibitem{James}
J. J. He, T.K. Ng, P. A. Lee, K. T. Law,
Phys. Rev. Lett. {\bf 112}, 037001 (2014).

\bibitem{Noah2}
N. F. Q. Yuan, Y. Lu, J. J. He, K. T. Law,
arXiv:1510.03137

\bibitem{Linder}
J. Linder and J. W. A. Robinson, 
Nat. Phys. {\bf 11}, 307 (2015).

\bibitem{Eschrig}
M. Eschrig, 
Phys. Today {\bf 64}, No. 1, 43 (2011).

\end{thebibliography}
\end{document}